\documentclass[final,twocolumn]{IEEEtran}

\usepackage{graphicx,epsfig,epstopdf}
\usepackage{amsmath}
\usepackage{color}
\usepackage{subfigure}
\usepackage{array}
\usepackage[noadjust]{cite}
\usepackage{multirow}
\usepackage[utf8]{inputenc}

\usepackage{cite}
\usepackage{etoolbox}

\begin{document}

	\title{Toward Ultimate Control of Terahertz Wave Absorption in Graphene}
	\author{Xuchen Wang,~\IEEEmembership{Student Member,~IEEE}, and Sergei A. Tretyakov,~\IEEEmembership{Fellow,~IEEE}  
	\thanks{This work was supported by the European Union's Horizon 2020 research and innovation programme-Future Emerging Topics (FETOPEN) under grant agreement No 736876.}
	\thanks{The authors are with the Department of Electronics and Nanoengineering, Aalto University, P.~O.~Box~15500, FI-00076 Aalto, Finland.}
	\thanks{Corresponding author: X. C. Wang, (email: xuchen.wang@aalto.fi).}}
	
	\maketitle

\begin{abstract}

It is commonly believed that weak light-matter interactions in low-mobility graphene dramatically limits  tunability of graphene-based optoelectronic devices, such as tunable absorbers or switches. 
In this paper, we  use a simple circuit model to understand absorption in graphene sheets. In particular, we show that light interacts weakly also with very high-mobility graphene sheets and propose systematic design means to overcome these problems. The results have allowed us to demonstrate  
in the terahertz band that perfect absorption with excellent electrical tunability can be achieved within a wide span of mobility values (from 200 to 20000~${\rm cm}^{2}{\rm V}^{-1}{\rm s}^{-1}$) which almost covers the whole range of  ever reported room-temperature mobilities. 	Remarkably, concentrating on the most practical low-mobility graphene devices, we exemplify our theory with two cases: frequency-tunable and switchable absorbers with near 100\% modulation efficiencies.
%
Our work provides  systematic and instructive insights into the design of highly tunable absorbers, without restrictions on graphene mobility. The design strategy and the developed analytical model can, in principle, be generalized to other wavelength regions from microwave to mid-infrared range, and other two-dimensional materials such as transition metal dichalcogenides (TMDs) and black phosphorus. 
\end{abstract}


\section{Introduction}

Graphene is an excellent platform for  optoelectronics \cite{bonaccorso2010graphene}, enabling better photodetectors, wave modulators \cite{sun2016optical,tamagnone2014fundamental} and tunable plasmonic devices \cite{yan2012tunable,vasic2013graphene}. The performance of graphene-based devices is largely determined by the intensity of light-matter interactions in graphene.
	It is well known that a pristine graphene layer naturally absorbs only a small portion of incident power in the full electromagnetic spectrum \cite{banman2015mid} (in the visible and near-infrared ranges it is $<$2.3\%, in mid-infrared, $<$3\% and  in far-infrared  it is $<$10\%). 
	In recent years, a lot of efforts have been made to enhance  absorption in graphene, targeting at different wavelength regions.
	In the visible and near-infrared ranges, absorption  can be improved using dielectric \cite{gan2012strong} or metallic resonators \cite{zhao2014enhancement}. In this range, electrical tuning the device is not possible since graphene exhibits universal conductance ($\sigma_{\rm s}={e^2}/{4\hbar}$) dominated by interband transitions.
	Probably, the most attractive spectrum for tunable graphene applications is from  mid-infrared to microwave band, where the conductivity of graphene can be dynamically controlled by electrostatic gating. According to the Kubo formula \cite{hanson2008dyadic}, the sheet impedance of graphene is inversely proportional to the  carrier mobility and the Fermi level.
	For a homogeneous graphene sheet, if the sheet impedance is engineered to the level of the free-space impedance $\eta_0=377~\Omega$ and the sheet is positioned in an appropriate resonant structure, the absorption in graphene will be strong. Based on this fact, most of the theoretical studies purposefully assume a large mobility or high Fermi level to make the sheet impedance of graphene close to $\eta_0$ \cite{pu2013strong,zhang2015tunable}. For example, under this assumption on mobility, graphene is patterned into different shapes to enable plasmon-induced absorption and realize  various functionalities. In particular, full absorption  \cite{thongrattanasiri2012complete,alaee2012perfect} as well as broadband \cite{khavasi2015design,ye2017broadband} and multi-band \cite{sun2018graphene,su2018tunable} absorption spectra have been reported.  

Despite the appeal of these theoretical predictions, it has been very difficult to realize this performance in laboratory.  Many researchers fabricate such nanopatterned graphene structures, but the observed absorptivity is much smaller than unity \cite{safaei2017dynamically,jang2014tunable}. 
The main reason of poor absorption is that the carrier mobility of processed graphene is much lower than those assumed in corresponding theoretical works. For example, in \cite{thongrattanasiri2012complete}, graphene mobility is assumed to be $\mu_{\rm m}=10000~{\rm cm}^{2}{\rm V}^{-1}{\rm s}^{-1}$. 
Such high mobility value is possible to obtain from mechanically exfoliated graphene \cite{petrone2012chemical}, but extremely hard from processed CVD graphene.  The typically reported mobility of CVD graphene is only around  $1000~{\rm cm}^{2}{\rm V}^{-1}{\rm s}^{-1}$ \cite{yao2013wide,jadidi2015tunable,ju2011graphene}, corresponding to sheet resistivity of $9000~\Omega$/sq (at a low Fermi level, $E_{\rm F}=0.1$~eV) which is severely mismatched with the free-space wave impedance.
The low carrier mobility of graphene is attributed to charged impurities and defects introduced during graphene growth, transfer and nanopatterning, which is unavoidable and difficult to control. 
Due to the hardship of obtaining high quality (high mobility) graphene, many experimental works resort  to improvements of the doping level of graphene, either with chemical or electrical methods, to compensate the poor carrier mobility \cite{balci2015graphene,kakenov2016observation,kim2018electronically}. For example, in \cite{kakenov2016observation}, full absorption of a graphene Salisbury screen occurs only when the Fermi level is raised up to $E_{\rm F}=1$~eV. 
As an alternative, in some works, graphene is subjected to chemical dopings by processing graphene in some chemical solutions such as ${\rm HNO_3}$ to increase its conductance \cite{yi2017tunable,min2014graphene}. Although in this way the graphene impedance can be engineered to the free-space impedance at ease, strong chemical doping may disrupt the band structure of graphene and thus compromize its tunability. Another method is to employ multilayer or sandwich-stacked graphene \cite{chen2018experimental,d2015terahertz,d2016adaptive,fallahi2012design} such that the effective impedance of graphene can be dramatically reduced to near free-space impedance if the numbers of layers are enough large.  In addition, graphene-metal hybrid structures are also utilized to reduce the effective impedance, and thus improve the modulation level \cite{tamagnone2014fundamental} and enrich the functionalities of graphene-based absorbers \cite{vasic2013graphene,huang2014graphene}.

While  the previous works rely on carefully tailoring the mobility value or the  Fermi level for perfect absorption, in this paper, we demonstrate that positioning a graphene sheet on a specific metasurface substrate, perfect and strongly tunable absorption is always possible even in very weakly-doped graphene, without  strict requirements of the carrier mobility. Understanding and design of such tunable graphene devices needs a clear physical model and systematic design strategy. 
We use a circuit model as an analytical tool and  show what factors  restrict absorption and tunability and  how to optimize performance for graphene with arbitrary mobility values. 
We demonstrate that, in the terahertz range, both  low and high carrier mobility weaken the absorption in graphene.  For graphene with low mobility, the large sheet resistance restricts  absorption, while for high quality graphene the kinetic inductance becomes a dominant decoupling factor. 
We show that both of these two decoupling problems can be solved by positioning graphene sheets on patch-type metasurfaces. We show that properly choosing the structural parameters enables perfect and strongly tunable absorption in graphene within a wide range of mobility values (from 200 to 20000~${\rm cm}^{2}{\rm V}^{-1}{\rm s}^{-1}$), even if the graphene sheet is very weakly doped. 
Particularly, in the low mobility regime, we design two tunable absorbers  with near 100\% modulation efficiencies, one has the tunable  frequency of absorption and the other operates as a switch between high absorption and high reflection regimes.

\section{Fundamental limits of absorption in continuous Graphene sheet}\label{section:limitation}

 The surface conductivity of graphene is generally spatial-temporal dispersive, $\sigma_{\rm g}(\textbf{\textit{k}}, \omega)$, where $\textbf{\textit{k}}$ and $\omega$ is the spatial and angular frequencies, respectively \cite{hanson2008dyadic}.  Since this work is in the scope of fast wave propagation (low-\textbf{\textit{k}}), the spatial dispersion (non-locality) of graphene conductivity can be neglected. The local surface conductivity is characterized by $\sigma_{\rm g}(\omega, T, E_{\rm F}, \gamma)$, where $T$ is the temperature, $E_{\rm F}$ is the Fermi level (or chemical potential) with respect to the charge neutrality point (CNP),  and $\gamma$ is the scattering rate. The local conductivity model is given by the well-known Kubo formula \cite{hanson2008dyadic},
\begin{equation}
	\begin{aligned}
		\sigma_{\rm g}(\omega)=\ 
		& -j\frac{e^2k_{\rm B}T}{\pi\hbar^2(\omega-j\gamma)}
		\left[\frac{E_{\rm F}}{k_{\rm B}T}+2\ln\left(1+e^{-\frac{E_{\rm F}}{k_{\rm B}T}}\right)\right]\\
		&-j\frac{e^2}{4\pi\hbar}\ln\left[\frac{2|E_{\rm F}|-(\omega-j\gamma)\hbar}{2|E_{\rm F}|+(\omega-j\gamma)\hbar}\right],
	\end{aligned}\label{Kubo}
\end{equation}
where $e$ is the elementary charge,  $k_{\rm B}$ is the Boltzmann constant, and $\hbar$ is the reduced Plank constant. The first term represents the intraband transitions contribution which is approximated under the condition of $\hbar\omega\ll2|E_{\rm F}|$, and the second term refers to the interband contribution. In the terahertz frequency band and at lower frequencies the intraband transitions dominate. Throughout this paper, we discuss applications in the terahertz band, approximating the surface conductivity of graphene with the first term in (\ref{Kubo}), but we note that the proposed design methodology is applicable in the general case. 

The Fermi energy is a function of the carrier density $n$, $E_{\rm F}=\hbar v_{\rm F}\sqrt{\pi n}$, which can be dynamically adjusted by applying a vertical electrostatic field upon graphene layer ($v_{\rm F}=10^8~{\rm cm/s}$ is the Fermi velocity). The scattering rate is inversely proportional to the Fermi energy, $\gamma={e{v_{\rm F}}^2}/({\mu_{\rm m} E_{\rm F}})$. $\mu_{\rm m}$ is the carrier mobility of graphene. In this work, we assume that the mobility is independent of the carrier concentration. 
The surface conductivity of graphene is a complex-valued parameter, and the graphene sheet can be considered as a complex-impedance sheet, with the sheet impedance  $Z_{\rm g}={1}/{\sigma_{\rm g}}=R_{\rm g}+jX_{\rm g}$.

\begin{figure}[h!]
	\centering
	\subfigure[]{
		\includegraphics[width=0.55\linewidth]{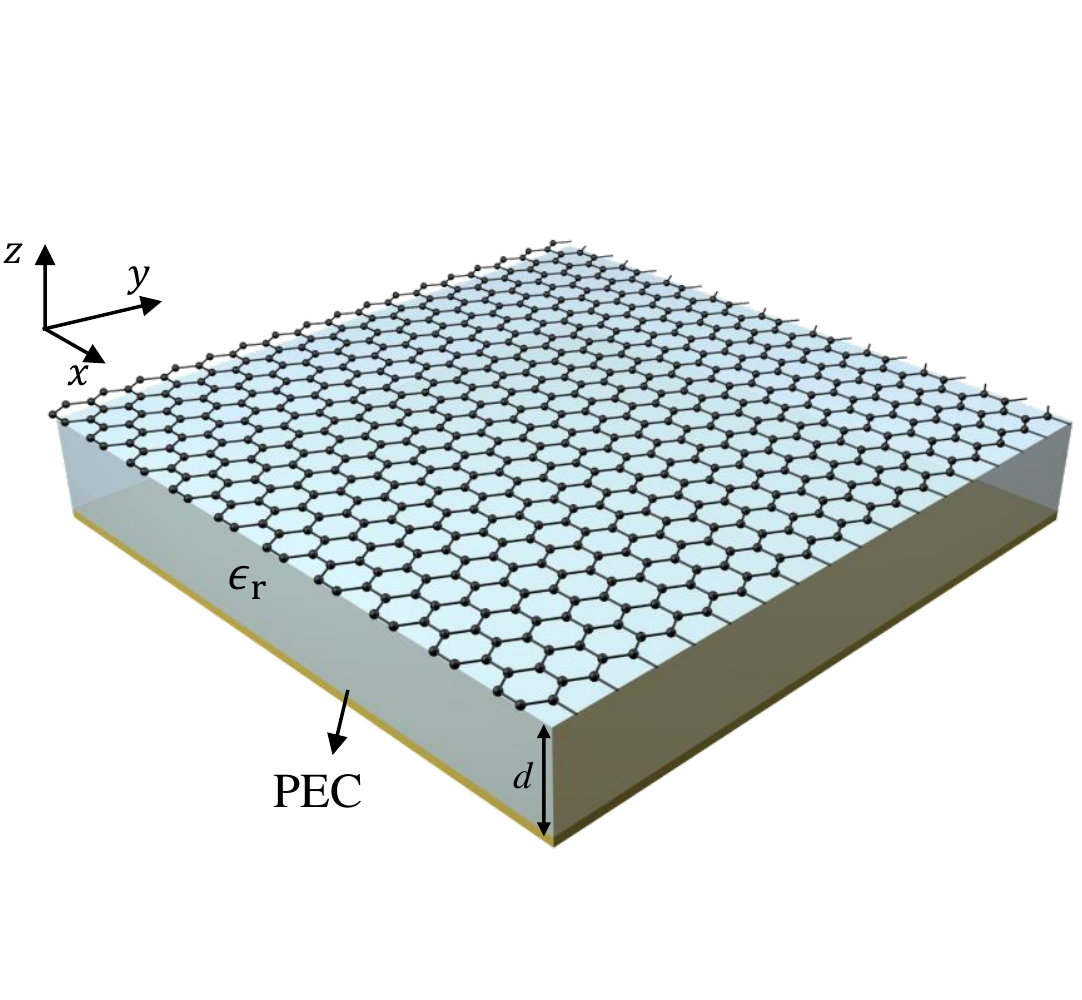}
		\label{fig:general_problem_structure}}
	\subfigure[]{
		\includegraphics[width=0.33\linewidth]{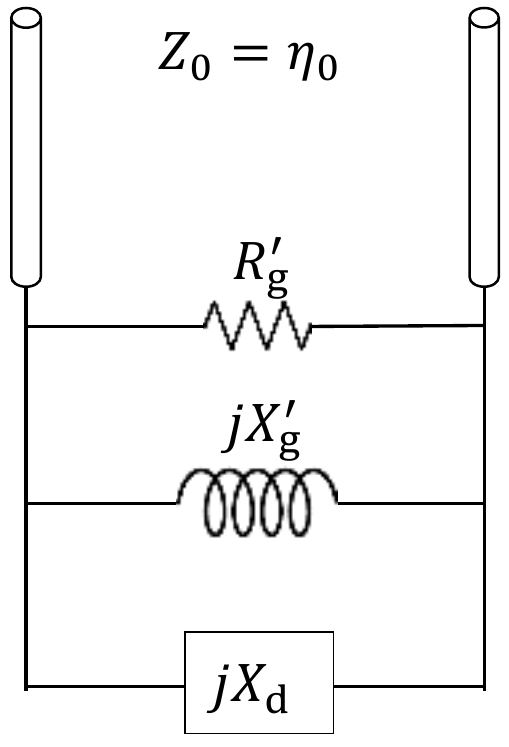}
		\label{fig:general_problem_circuit}}
	\caption{(a) Schematic and  (b) circuit model of graphene-based Salisbury screen.}\label{general_problem_structure_circuit}
\end{figure}

Probably the simplest arrangement which allows total absorption in graphene is the Salisbury screen \cite{absorbers_review,kakenov2016observation,min2014graphene}, where the transmission is totally blocked.  The classical Salisbury screen is a resistive sheet placed on a grounded dielectric substrate of the quarter wavelength thickness. Total absorption is realizable if the input impedance of the structure can be made equal to the free-space impedance $Z_0$. Here, a graphene layer is used as a resistive sheet, as shown in Fig.~\ref{fig:general_problem_structure}. It is noteworthy that the thickness of the substrate is not necessary equal to one quarter of the wavelength since the graphene is not a purely resistive layer and possesses some reactive properties. 
The very effective way to analyze graphene-based metasurfaces is the transmission-line model. The corresponding equivalent circuit can clearly manifest the physical mechanism of the devices and even  analytically evaluate their performances, with either local \cite{jadidi2015tunable,wang2015reconfigurable,khavasi2015design} or nonlocal \cite{correas2013spatially,gomez2013effect} graphene conductivity. Here,  we also analyze this structure with a simple transmission-line model shown in Fig.~\ref{fig:general_problem_circuit}.
Instead of writing $Z_{\rm g}$ as a series connection of a resistor and a reactive element, as many other authors do \cite{jadidi2015tunable,wang2015reconfigurable,khavasi2015design}, we express the graphene sheet impedance as two parallel components, the shunt resistance $R_{\rm g}^\prime$ and the shunt reactance $jX_{\rm g}^\prime$ [see Fig.~\ref{fig:general_problem_circuit}]. 
Obviously, $R_{\rm g}^\prime$ and $jX_{\rm g}^\prime$ can be expressed in terms of the series-circuit parameters   $R_{\rm g}$ and $X_{\rm g}$: \begin{equation}
	R_{\rm g}^\prime=R_{\rm g}+\frac{X_{\rm g}^2}{R_{\rm g}}
	\label{shunt_Rg}
\end{equation}
and
\begin{equation}
	jX_{\rm g}^\prime=j\left(X_{\rm g}+\frac{R_{\rm g}^2}{X_{\rm g}}\right).\label{shunt_Xg}
\end{equation}
Using this convenient representation, one can clearly understand the physical mechanism of graphene absorption and apply impedance matching method to control the absorption, as can be seen in the following discussions.

The grounded substrate is represented by a shorted transmission line with the characteristic impedance $\eta_{\rm d}=\sqrt{\mu_{\rm 0}/\epsilon_{\rm 0}\epsilon_{\rm r}}$. For normal incidence of plane waves (except Section \ref{oblique}, all the discussions and results in this paper correspond to normal incidence scenario), the shorted transmission line can be viewed as a reactance whose impedance is $jX_{\rm d}=j\eta_{\rm d}\tan(k_{\rm d}d)$,
where  $k_{\rm d}=\omega\sqrt{\epsilon_{\rm r}\epsilon_{\rm 0}\mu_{\rm 0}}$ is the wave number in the substrate.
The total input admittance of the structure can be written as $	Y_{\rm in}={R_{\rm g}^\prime}^{-1}+(jX_{\rm g}^\prime)^{-1}+(jX_{\rm d})^{-1}$.
Perfect absorption takes place when $Y_{\rm in}=Y_{\rm 0}={1}/{Z_0}$, which satisfies the impedance matching condition. Thus, the real and imaginary parts of $Y_{\rm in}$ should satisfy
\begin{equation}
	\frac{1}{R_{\rm g}^\prime}=\frac{1}{Z_0}\label{real_condition}
\end{equation}
and
\begin{equation}
	\frac{1}{jX_{\rm g}^\prime}+\frac{1}{jX_{\rm d}}=0.\label{imag_condition}
\end{equation}
Condition (\ref{real_condition}) means that the shunt resistance of the graphene sheet should be equal to the free-space impedance at the frequency of perfect absorption (for normal incidence, $Z_0=\eta_0=\sqrt{\mu_0/\epsilon_0}$).
The difference between $R_{\rm g}^\prime$ and $Z_0$ actually determines the absorption level.
Equation~(\ref{imag_condition}) tells that the reactances of the graphene sheet and the input reactance of the grounded substrate should cancel out (forming a high-impedance surface). This condition is usually easier to satisfy [compared to  (\ref{real_condition})] because it can be easily met by adjusting the substrate thickness or its dielectric constant, for an arbitrary value of $X^\prime_{\rm g}$. 
Assuming that the condition (\ref{imag_condition}) is satisfied, the reflection  coefficient is written as $r=(R_{\rm g}^\prime-Z_0)/(R_{\rm g}^\prime+Z_0)$.
We see that the reflectance of the structure is completely determined by the shunt resistance of graphene. If $R_{\rm g}^\prime$ deviates from $Z_0$, the wave interacts weakly with graphene and the absorption decreases. 
	Here, we define the logarithm of the normalized shunt resistance as a decoupling factor, $D_e=\log (R_{\rm g}^\prime/Z_0)$. The decoupling factor measures the extent of impedance mismatch: wave can be fully absorbed in graphene only if $D_e=0$; large absolute values of $D_e$ correspond to low absorption.

\begin{figure}[h!]
	\centering
	\includegraphics[width=0.85\linewidth]{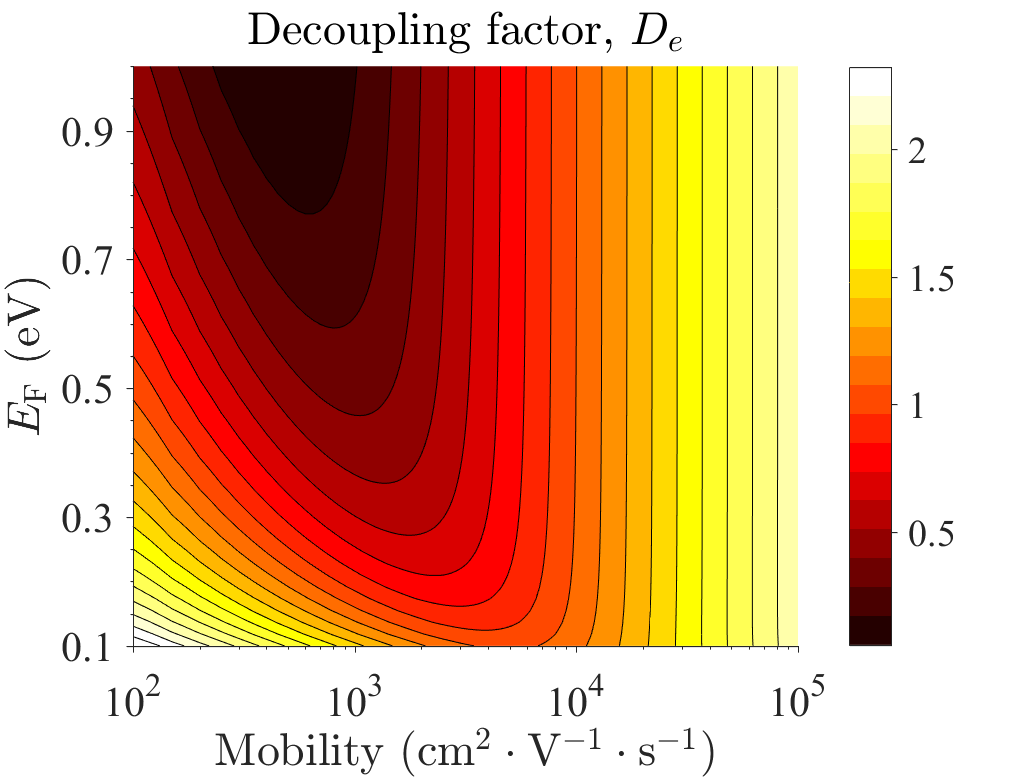}
	\caption{Decoupling factor $D_e$ as a function  of the carrier mobility and the Fermi level at 4~THz.  }\label{fig:shunt_resistance_Mob_EF_5T}
\end{figure}

Next, in the terahertz region, we study the decoupling mechanisms for graphene with arbitrary mobility and Fermi levels. 
	Figure~\ref{fig:shunt_resistance_Mob_EF_5T} shows the calculated $D_e$ as a function of graphene mobility and Fermi level at 4~THz. We study the doping range from 0.1~eV to 1~eV, which can be realized  with high capacitance ion gel films \cite{safaei2017dynamically,kakenov2016observation,hu2015broadly}.   
	For low quality graphene ($\mu_{\rm m}<2000~{\rm cm}^{2}{\rm V}^{-1}{\rm s}^{-1}$), we can see that   decoupling is strong and absorption is very weak at low Fermi levels. This is caused by the huge sheet resistance of weakly-doped graphene ($R_{\rm g}^\prime$ is dominated by $R_{\rm g}$). Therefore, improving the doping level (decreasing $R_{\rm g}$) will somewhat enhance the absorption. 
	This can clearly explain the recent experimental observation of perfect terahertz absorption in graphene \cite{kakenov2016observation}, where graphene is subjected to a high electrical doping ($E_{\rm F}=1$ eV).
	For high quality graphene, the impedance mismatch becomes very severe and it cannot be alleviated by the increase of Fermi level. The decoupling is dominated by the kinetic inductance of graphene ($R_{\rm g}^\prime$ is mostly affected by $X_{\rm g}$). This theory can successfully explain the absorption enhancement in graphene plasmonic patterns \cite{thongrattanasiri2012complete}. In fact, the capacitive interactions between the neighboring graphene cells cancel the natural kinetic inductance of graphene, thus increasing the coupling.
	We see that at low doping levels, both low and high quality graphene sheets are difficult to couple with incident waves, although   decoupling is caused by different mechanisms (large sheet resistance or reactance).
	Hereafter, we aim at perfect absorption in weakly-doped graphene. 
	
\begin{figure}[h!]
	\centering
	\includegraphics[width=0.85\linewidth]{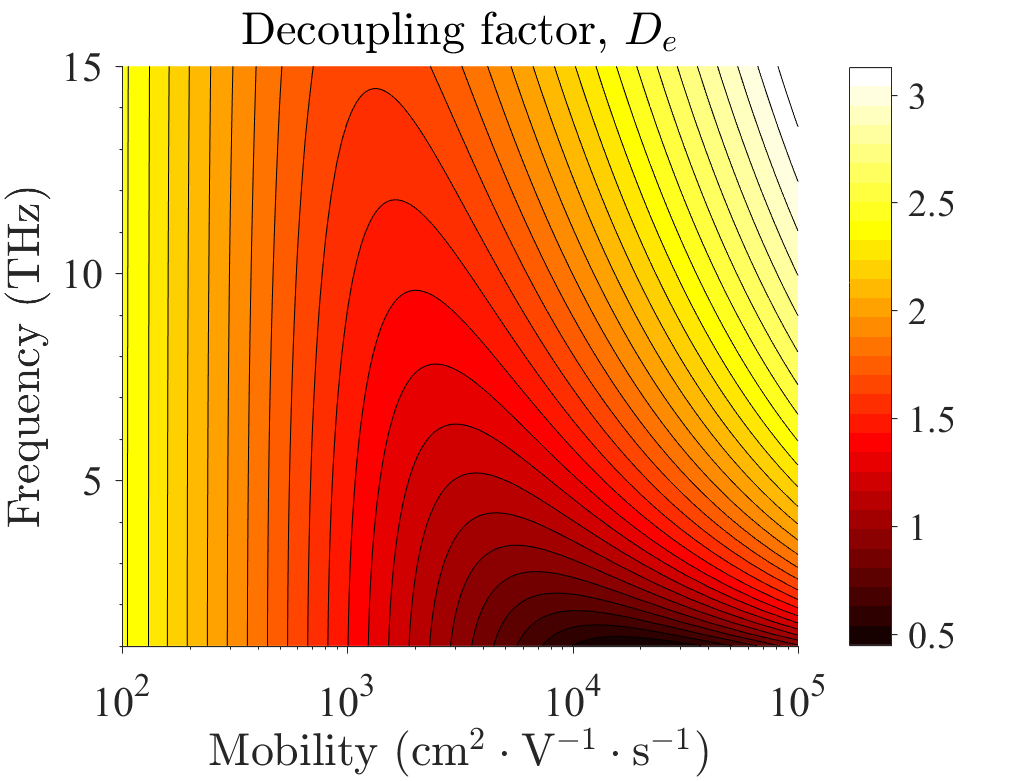}
	\label{fig:parallel_real_0.1eV}
	\caption{Decoupling factor  as a function  of the carrier mobility and the frequency at  $E_{\rm F0}=0.1$~eV.  }\label{Parallel_impedance}
\end{figure}

It is known that a pristine graphene layer always becomes doped by some residual carriers during the growth and processing of graphene. This initial doping is unintentional, and varies from sample to sample.  
		We take a typical value of this unintentional doping as $n_0=7.3\times10^{11}{\rm cm}^{-2}$, corresponding to the Fermi level of $E_{\rm F0}=0.1$~eV \cite{tan2007measurement,dabidian2016experimental}.
		Realizing perfect absorption with such low Fermi levels is more difficult than in heavily-doped graphene, but beneficially, it may enable the use of graphene in perfect absorbers without external assists (electrical or chemical doping).
		More importantly, if the matching condition is satisfied in a weakly-doped graphene, it is possible to maximize the tunability because one can drive the Fermi level from a very low state to the practical upper limit, which provides a maximum force to break the established matching. 
		Figure~\ref{Parallel_impedance} shows the decoupling factor for $E_{\rm F}=0.1$~eV at different frequencies and for different mobility values.
		In the high mobility regime, the inductance-induced decoupling becomes evident as the frequency increases ($X_{\rm g}$ increases fastly and becomes much larger than $R_{\rm g}$), while for low quality graphene it is almost frequency-independent ($R_{\rm g}$ is much larger than $X_{\rm g}$). 
		From Fig.~\ref{Parallel_impedance}, we see that strong decoupling is a universal problem in the whole terahertz range for weakly doped graphene.  
		In the next section,  we will demonstrate that strong but differently induced decoupling  can be eliminated by using  one type of metasurface substrates, and perfect absorption is always achievable in graphene irrespective of the carrier mobility.



\section{Perfect absorption in graphene using metallic metasurface substrate}


\subsection{ Analytical modeling } \label{Theory}
\begin{figure}[h!]
	\centering
	\subfigure[]{
		\includegraphics[width=0.6\linewidth]{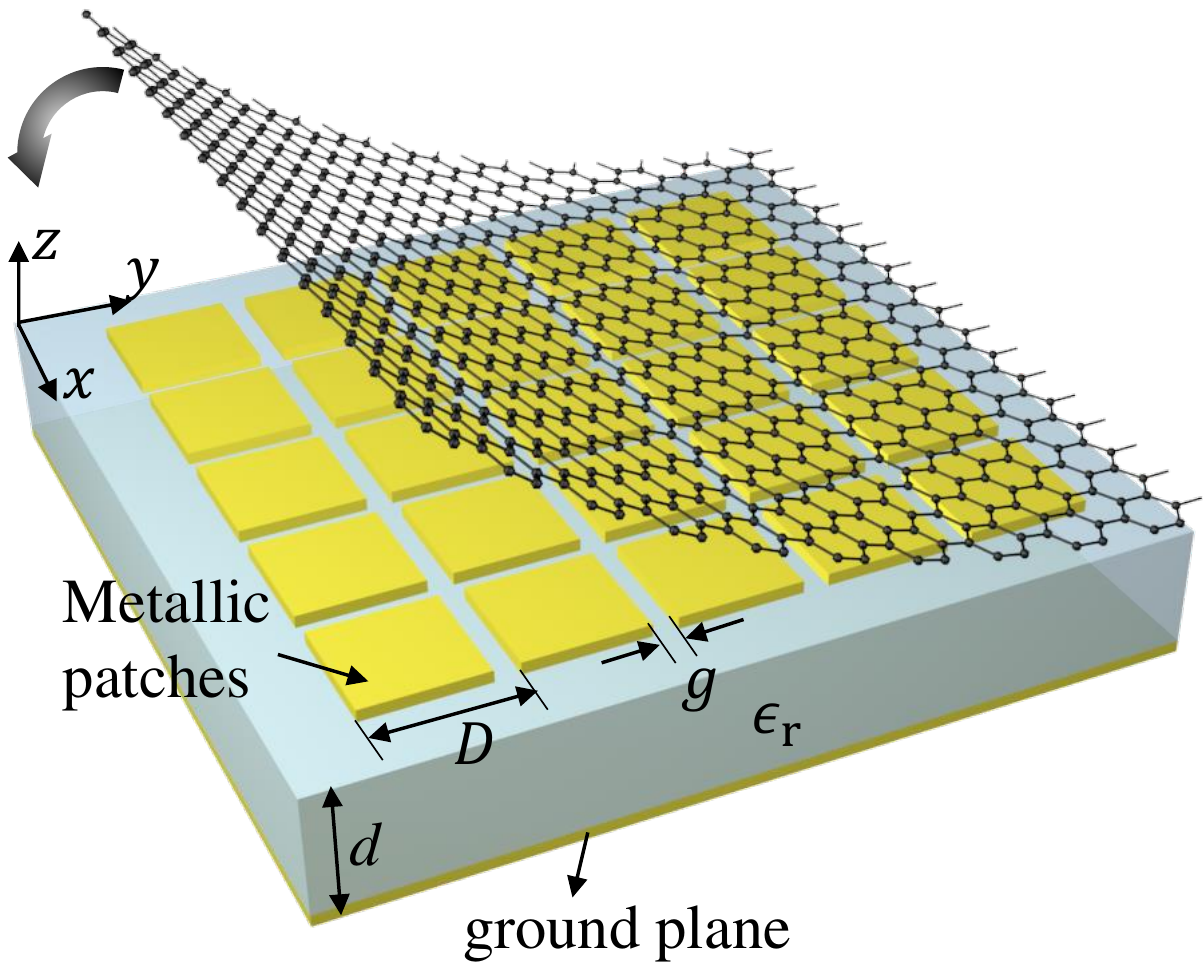}
		\label{fig:proposed_structure}}
	\subfigure[]{
		\includegraphics[width=0.31\linewidth]{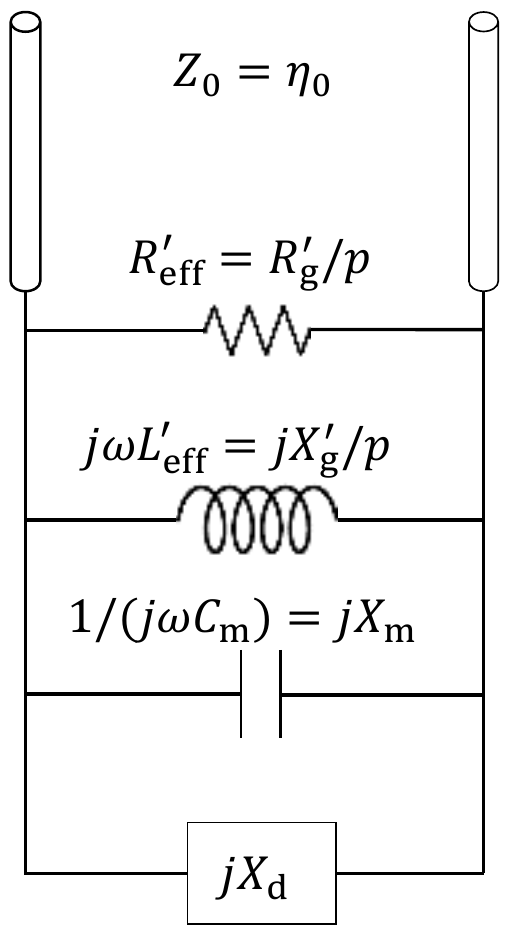}
		\label{fig:proposed_circuit_model}}
	\caption{(a) Schematics of the proposed graphene absorber based on metasurface substrate  (b) equivalent circuit model of the proposed structure for normal incidence.}\label{proposed_configuration}
\end{figure}

The schematic of the metasurface-based graphene absorber is displayed in Fig.~\ref{fig:proposed_structure}. Compared to the Salisbury structure,  we use one auxiliary layer  formed by thin metal patches periodically positioned on the substrate. The patch array period is small compared to the wavelength, so that there are no propagating Floquet modes. The graphene sheet is  directly transferred on the patterned metal layer.
The metasurface (formed by the periodic square patch array and the grounded substrate) plays two important roles: creation of a capacitive component, together with inductive graphene to form a  resonant cavity (high impedance surface), and  reducing the effective shunt resistance of graphene.

\begin{figure}[h!]
	\centering
	\subfigure[]{
		\includegraphics[width=0.465\linewidth]{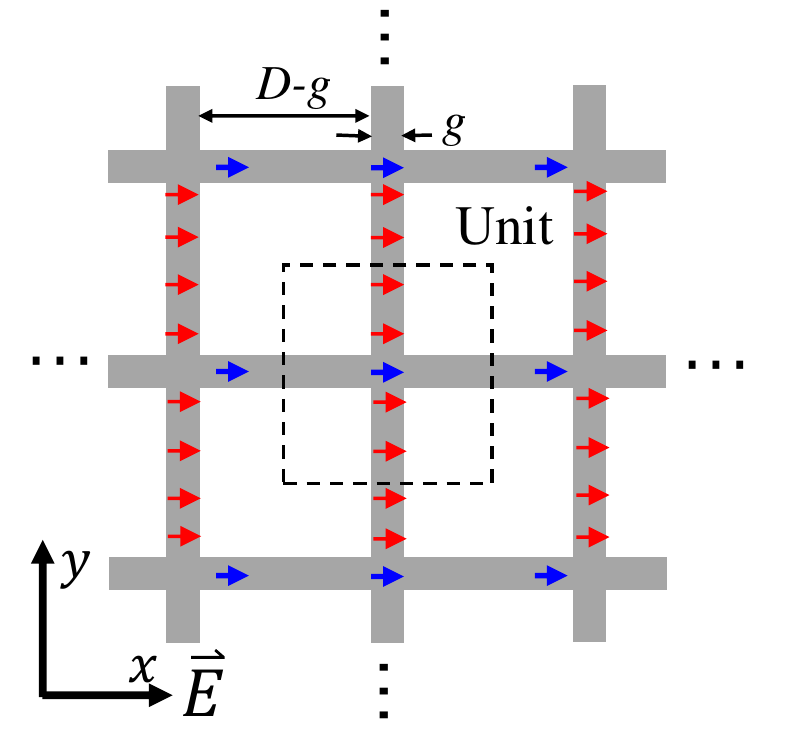}
		\label{fig:mesh}}
	\subfigure[]{
		\includegraphics[width=0.465\linewidth]{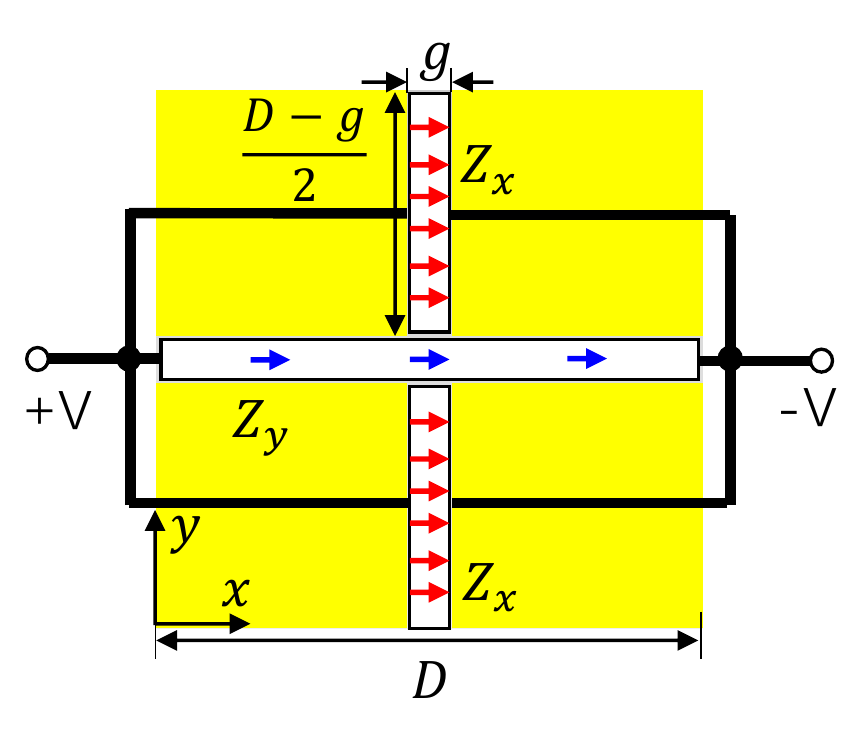}
		\label{fig:circuit}}
	\caption{(a) Top view of the \textquotedblleft patterned\textquotedblright~graphene. The red arrows represent strong induced currents in the $y$-oriented strips due to capacitive coupling between the patches, and the blue arrows represent weak currents in the $x$-oriented strips. The black dashed region represents one unit cell.  (b) Circuit analogue of graphene sheet in one unit cell. The yellow region is covered by metal, and white strips are open graphene.  }\label{fig:complementary_structure}
\end{figure}

With the presence of metallic patches, the graphene layer is equivalent to a \textquotedblleft patterned fishnet\textquotedblright~structure (a complementary structure to the array of square patches), as shown in Fig.~\ref{fig:mesh}. This is because in the graphene-metal contact areas graphene is effectively shorted by highly conductive metal. Tangential electric field on graphene  surface where it is supported by metal is zero due to the boundary condition on metal. Thus,  the incident wave is not able to perceive the graphene sheet on metal and there is no current induced in this part of the graphene layer. The only effective part is the graphene \textquotedblleft fishnet\textquotedblright~which is not shorted.  
Assuming that the electric field is polarized  along the $x$-direction, strong capacitive coupling  between the patches occurs via $y$-directed gaps. Therefore,  in the effectively \textquotedblleft patterned\textquotedblright~graphene only $y$-oriented strips are strongly excited [see the red arrows in Fig. \ref{fig:mesh}]. Figure~\ref{fig:circuit} illustrates the circuit analogue of one unit cell. The $x$-polarized electric field is viewed as a voltage source. The graphene is \textquotedblleft patterned\textquotedblright~into three parallel-connected impedance strips with impedances $Z_x$, $Z_y$, and  $Z_x$. The impedances of the three strips can be calculated as their length-width ratios multiplied by   the intrinsic sheet impedance of graphene $Z_{\rm g}$ (the length refers to the edge along the current flowing direction):
\begin{equation}
Z_x=\frac{2g}{D-g}Z_{\rm g}, \quad Z_y=\frac{D}{g}Z_{\rm g},
\end{equation} 
where $D$ and $g$ are the cell period and the gap width, respectively. 
Under the condition of $D\gg g$ (this assumption  holds throughout the paper), $Z_y$ is very large and it can be viewed as an open circuit, which in turn explains the weak currents in the $x$-directed strip. Thus, the total effective admittance of graphene can be approximated as
\begin{equation}
\frac{1}{Z_{\rm eff}}=\frac{2}{Z_x}+\frac{1}{Z_y}\approx\frac{2}{Z_x}. \label{effective impedance of graphene}
\end{equation} 
From (\ref{effective impedance of graphene}) we see that the effective impedance of graphene in one cell can be written as $Z_{\rm eff}=Z_{\rm g}/p$, where $p=(D-g)/g$ will be called the \textit{scaling factor.} Apparently, if $p$ is enough large, the effective impedance of unshorted graphene will be greatly reduced, which is necessary for the enhancement of absorption.

The equivalent circuit of the proposed structure is shown in Fig.~\ref{fig:proposed_circuit_model}. It is modified from the circuit of the Salisbury screen [see Fig.~\ref{fig:general_problem_circuit}] by scaling the graphene impedance down to $Z_{\rm g}/p$, as well as adding a parallel capacitive component contributed by the periodically arranged metal patches. In this circuit model, the metal is treated as perfect conductor. This is a good approximation in the low terahertz band when metal is gold or silver. The grid impedance of the patch array is capacitive, $jX_{\rm m}={1}/{j\omega C_{\rm m}}$.
For a normally incident plane wave, $C_{\rm m}$ reads
\begin{equation}
	C_{\rm m}=\frac{2\epsilon_{\rm eff}\epsilon_0D\ln\left(\csc\frac{\pi}{2(p+1)}\right)}{\pi}, \label{Cm_a}
\end{equation}
where $\epsilon_{\rm eff}=(\epsilon_{\rm r}+1)/2$ \cite{luukkonen2008simple}.

Similarly to the Salisbury screen discussed in the previous section, perfect absorption takes place when the following two conditions are satisfied:
\begin{equation}
	\frac{R^\prime_{\rm g}}{p}=Z_0\label{matching:real}
\end{equation}
and 
\begin{equation}
	\frac{1}{jX^\prime_{\rm g}/p}+\frac{1}{jX_{\rm m}}+\frac{1}{jX_{\rm d}}=0.\label{condition: high impedance surface}
\end{equation}
In what follows, the design procedures for unity absorption at a specified frequency  $f_0$ are introduced.
For a given graphene sample with arbitrary mobility $\mu_{\rm m}$ and initial Fermi-level $E_{\rm F0}$, we can obtain the sheet impedance in its shunt form using Equation (\ref{Kubo})-(\ref{shunt_Xg}). Then $p$  is determined from condition (\ref{matching:real}). 
For a fixed graphene sample and substrate, $jX_{\rm d}$ and $jX^\prime_{\rm g}$ are already known. Thus,  $D$ can be uniquely solved to satisfy  condition (\ref{condition: high impedance surface}). 
However, the obtained $D$ may be much larger than the substrate thickness $d$.
In this case the analytical formula for the grid capacitance (\ref{Cm_a})  becomes not accurate  due to ignorance of evanescent-mode coupling with the ground plane \cite{tretyakov2003dynamic}. 
In this case, one should use more accurate expressions of $C_{\rm m}$  \cite{tretyakov2003dynamic} where the influence of evanescent modes is accounted for.
In the following validation examples we set up a relation between the thickness of the substrate and the unit-cell size as $d=aD$, where $a$ is a parameter larger than $0.3$. In this case, the analytical formula (\ref{Cm_a}) is always valid \cite{costa2013circuit}. 
With condition (\ref{condition: high impedance surface}), the patch-array period can then be determined.

After $p$, $D$ and $g$ have been determined, we can analytically calculate the absorption spectrum. The total input impedance of the structure is 
\begin{equation}
	Z_{\rm in}=\frac{Z_{\rm g}X_{\rm m}X_{\rm d}}{pX_{\rm m}X_{\rm d}-jZ_{\rm g}(X_{\rm m}+X_{\rm d})}\label{input impedace}.
\end{equation}
The reflection and absorption coefficients read 
\begin{subequations}
		\begin{align}
		R=\frac{Z_{\rm in}-Z_0}{Z_{\rm in}+Z_0},\\
		A=1-|R|^2.
		\end{align}\label{reflection coefficient}
\end{subequations}

\subsection{Generalization for oblique-incidence scenarios} \label{oblique}
In the above analysis, we consider only the normal-incidence scenario. In this case, due to the $\pi/2$ rotational symmetry of the metasurface substrate, the structure responses identically for waves polarized along the $x$ and $y$ directions. However, for oblique illuminations the scattering properties become dependent on both the incident angle and the polarization state.  Without loss of generality, the circuit model in Fig.~\ref{fig:proposed_circuit_model} used  for the normal incidence can be applied in oblique-incidence scenarios by making only the following modifications of the equivalent circuit parameters: 
\begin{enumerate}
\item For TE polarization, the equivalent capacitance of patch array is affected by the incident angle $\theta_{\rm i}$, while it is stable for TM incidences~\cite{luukkonen2008simple},
	\begin{subequations}
		\begin{align}
		&C_{\rm m}^{\rm TE}(\theta_{\rm i})=C_{\rm m}\left(1-\frac{\sin^2\theta_{\rm i}}{\epsilon_{\rm r}+1} \right),\\
		&C_{\rm m}^{\rm TM}(\theta_{\rm i})=C_{\rm m}.
		\end{align}\label{patch impedance}
	\end{subequations}
\item The input reactance of the grounded substrate is also a function of the incident angle for both polarizations \cite{luukkonen2008simple}, and it reads
	\begin{subequations}
	\begin{align}
	jX_{\rm d}^{\rm TE}(\theta_{\rm i})=\frac{j\eta_{\rm d}\tan(k_{\rm n} d)}{\sqrt{1-\frac{\sin^2\theta_{\rm i}}{\epsilon_{\rm r}}}},\qquad \quad\\
	jX_{\rm d}^{\rm TM}(\theta_{\rm i})={j\eta_{\rm d}\sqrt{1-\frac{\sin^2\theta_{\rm i}}{\epsilon_{\rm r}}}\tan(k_{\rm n} d)}.
	\end{align}\label{substrate impedance}%
\end{subequations}
Here, $k_{\rm n}=\sqrt{k_0^2-k_{\rm t}^2}$ is the normal component of the  wavevector in the substrate with the transverse wavenumber  $k_{\rm t}=k_0\sin\theta_{\rm i}$.
\item The wave impedances in free space change to
	\begin{equation}
	Z_0^{\rm TE}(\theta_{\rm i})=\frac{\eta_0}{\cos\theta_{\rm i}},\quad  Z_0^{\rm TM}(\theta_{\rm i})=\eta_0\cos\theta_{\rm i}.\label{wave impedance}
	\end{equation}
\end{enumerate}
It should be noted that the sheet impedance of graphene does not change with the incident angle and polarization states, since spatial dispersion in this layer is negligible. With those modifications of  the circuit model, one can then follow the method of impedance matching and replace $Z_0$, $X_{\rm m}$ and $X_{\rm d}$  in (\ref{matching:real}) and (\ref{condition: high impedance surface}) with their angle-dependent generalized expressions listed in (\ref{patch impedance}), (\ref{substrate impedance}) and (\ref{wave impedance}). From ~(\ref{wave impedance}), we can see that the TE wave impedance increases with growing incident angle. For grazing  incident angles, $Z_0^{\rm TE}(\theta_{\rm i})$ can be comparable to the large shunt resistance of graphene. In this case, perfect absorption may be possible even without the assistance of metallic patches. This explains the enhanced absorption of graphene Salisbury screen at TE grazing incidences reported in \cite{zhu2016angle,ying2017enhanced}.
 
From (\ref{patch impedance}), (\ref{substrate impedance}), and (\ref{wave impedance}) we can see that both the free-space impedance and the input impedance of the metasurface substrate change with $\theta_{\rm i}$. Therefore, if perfect absorption is defined at normal incidence, the increase of the incident angle will  degrade the absorption level since the matching conditions (\ref{matching:real}) and (\ref{condition: high impedance surface}) are not satisfied anymore. 
Equation~(\ref{patch impedance}) and (\ref{substrate impedance}) indicate that with a high dielectric constant, the patch capacitance (for TE wave) and the substrate reactance are insensitive to the incident angle, which can provide angle-insensitive metasurface substrates \cite{luukkonen2008simple}. In this case one can tune the shunt resistance of graphene to match the altered free-space impedance. In this way, high absorption is possible for a wide range of incident angles. Another method is to embed periodically arranged metal pins in the dielectric substrate to form a mushroom-type metasurface, which can stabilize the substrate reactance but only for TM polarized waves \cite{maslovski2003}.  With this method it is possible to realize perfect absorption for TM waves at all incident angles (from normal to grazing incidence) by tuning the resistance of graphene, as demonstrated in our recent work \cite{wang2018graphene}.

\subsection {Perfect absorption in weakly-doped graphene}

\begin{figure}[h!]
	\centering
	\includegraphics[width=0.9\linewidth]{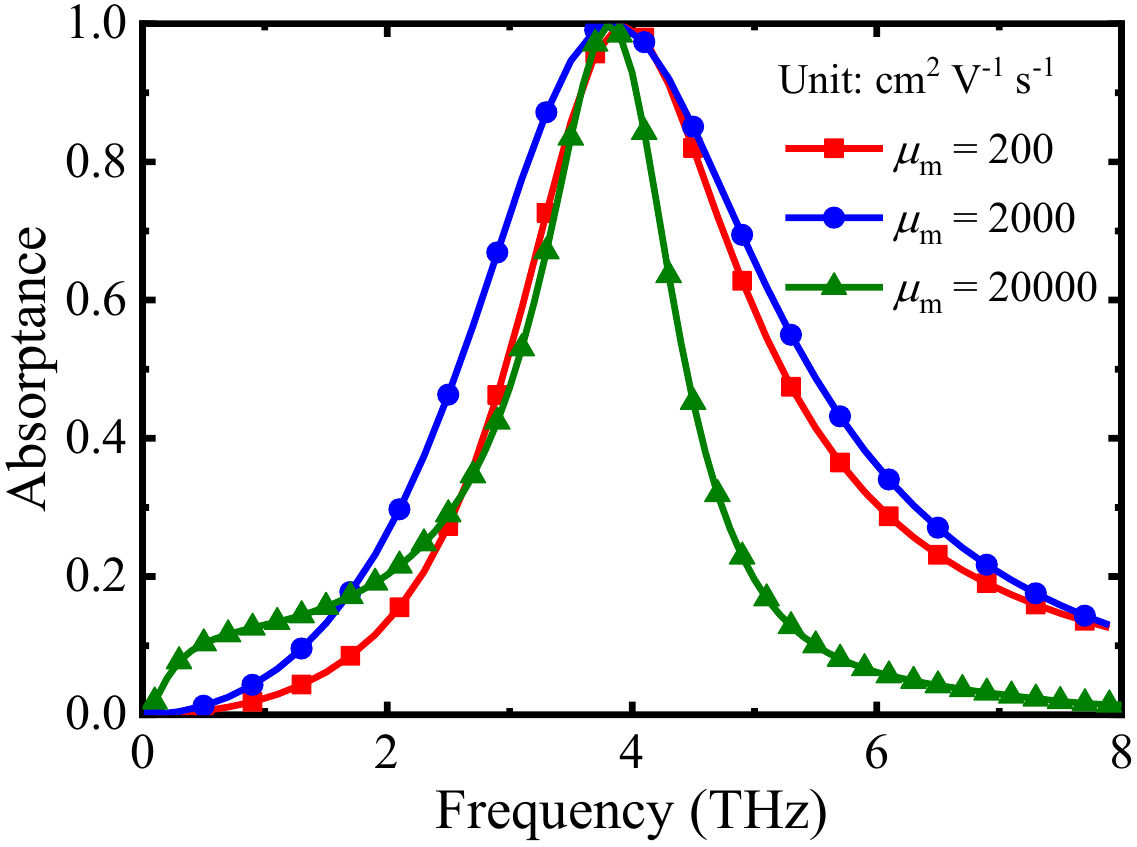}
	\caption{Simulated absorption coefficient in terms of frequency for different qualities of graphene. The structure dimensions are analytical solutions. $D$ is 7~$\mu$m, 10~$\mu$m, 17~$\mu$m, and $p$ is 136, 15.9, and 24.8 for graphene mobility $\mu_{\rm m}=$~200, 2000, and 20000~${\rm cm}^{2}{\rm V}^{-1}{\rm s}^{-1}$ (their scattering time is $\tau=1/\gamma=$~ 2~fs, 20~fs and 200~fs), respectively. The substrate thickness is set to be ratio $d=0.5D$. The dielectric constant of the substrate is assumed to be $\epsilon_{\rm r}=2.8$.}\label{fig:differnt mobilities}
\end{figure}

Hereafter, we will focus on the scenario of normal incidence. Following the design procedures elaborated in Section \ref{Theory},  we can achieve perfect absorption in graphene of different quality (within a wide span of carrier mobilities). Here, we investigate graphene mobilities from $\mu_{\rm m}=200~{\rm cm}^{2}{\rm V}^{-1}{\rm s}^{-1}$ to $\mu_{\rm m}=20000~{\rm cm}^{2}{\rm V}^{-1}{\rm s}^{-1}$. 
The initial doping of graphene is assumed to be $E_{\rm F0}=0.1$~eV.
The example target frequency for unity absorption is 4~THz.  The required $D$ and $p$ for each graphene mobility value  are solved from conditions (\ref{matching:real}) and (\ref{condition: high impedance surface}). Numerical tool HFSS (based on the finite elements method) is employed to simulate the proposed structure. Graphene is modeled as a zero-thickness sheet, with the frequency-dependent sheet impedance given by 
(\ref{Kubo}). We use a 75~nm-thick gold layer for the metallic ground plane and the patches. Gold conductivity is modeled by the Drude formula with the parameters $\gamma_{\rm Au}=3.33\times10^{13}$~rad/s and $\omega_{\rm p}=1.36\times10^{16}$~rad/s \cite{jadidi2015tunable}.  Figure~\ref{fig:differnt mobilities} displays the simulated absorption coefficients for each assumed mobility, using the analytically solved structural parameters of metasurface. All of the samples have nearly perfect absorption at the designated frequency of 4~THz. 

\begin{figure}[h!]
	\centering
	\includegraphics[width=0.9\linewidth]{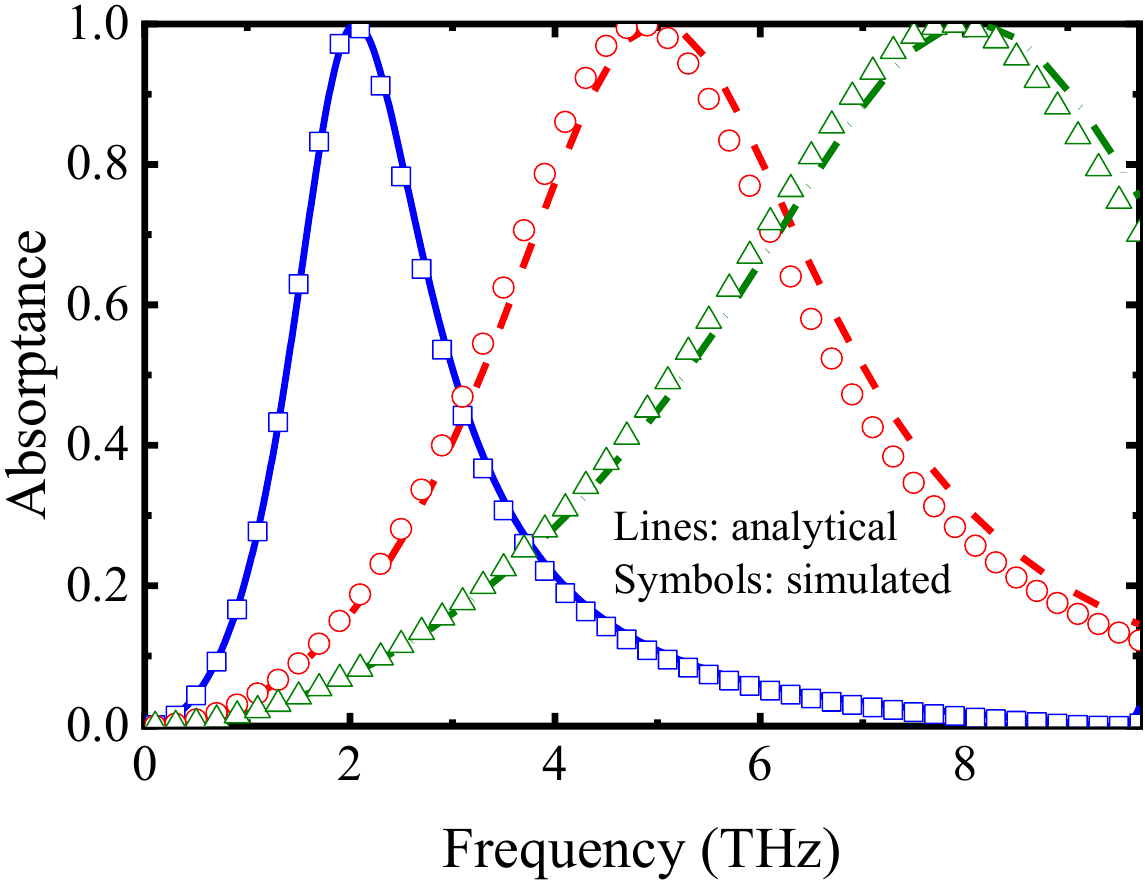}
	\caption{Analytical and numerical results for total absorption designed at different frequencies. $D$ is calculated to be 19 $\mu$m, $8\mu$m and $5\mu$m for $f=2$~THz, 5~THz and 8~THz, respectively.  The substrate thickness is set to be ratio $d=0.5D$. The dielectric constant of the substrate is assumed to be $\epsilon_{\rm r}=2.8$. The Fermi level is set to be 0.1~eV and the mobility is fixed as $\mu_{\rm m}=$~2000~${\rm cm}^{2}{\rm V}^{-1}{\rm s}^{-1}$ ($\tau=20$~fs at 0.1~eV). }\label{fig:analytical_simulation_comp}
\end{figure}

The proposed method is also suitable for designing perfect absorbers at different frequencies while using graphene of the same quality.
Here we use a low-mobility graphene with $\mu_{\rm m}=2000~{\rm cm}^{2}{\rm V}^{-1}{\rm s}^{-1}$ and the target absorption frequencies are 2~THz, 5~THz, and 8~THz with required scaling factors $p=14.4$, 17.2 and 22.9.
The analytical absorption spectra are calculated  using (\ref{input impedace}) and (\ref{reflection coefficient}), as shown in Fig.~\ref{fig:analytical_simulation_comp} (lines). With the same structural parameters, the simulated absorptances (symbols) are compared with the analytical ones in Fig.~\ref{fig:analytical_simulation_comp}.
Good agreement between analytical and numerical absorptance in the frequency range confirms the accuracy of the developed analytical model. 
It should be noted that the analytical formulas based on the circuit model may not be valid in the near-infrared and visible ranges since the metal dispersion also needs to be considered in these ranges. In addition,  the proposed topology  allows to ensure perfect absorption (impedance matching) only at one specific frequency point. Deviation from the targeted frequency degrades the level of absorption due to the dispersion of graphene and metasurface substrate. In this scenario, the matching conditions of (\ref{matching:real}) and (\ref{condition: high impedance surface}) will not hold. In the next section, we will discuss how to realize impedance matching in a continuous spectrum via electrical tuning of graphene. 

\section{Electrical tunability}\label{tunability}

\begin{figure*}
	\centering
	\subfigure[]{
		\includegraphics[width=0.3\linewidth]{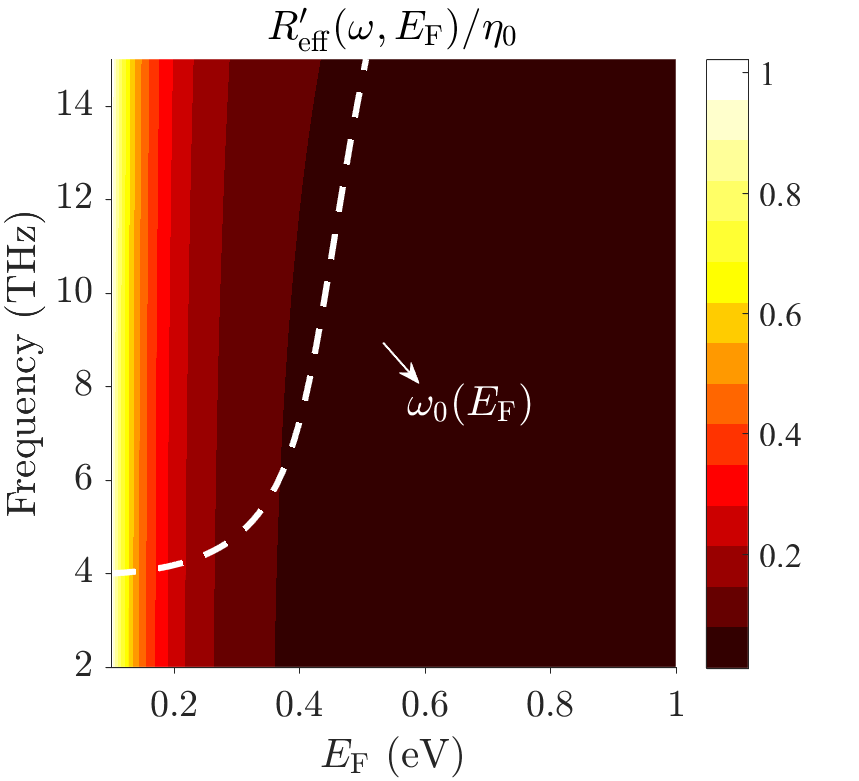}
		\label{fig:200_mue_freq}}
	\subfigure[]{
		\includegraphics[width=0.3\linewidth]{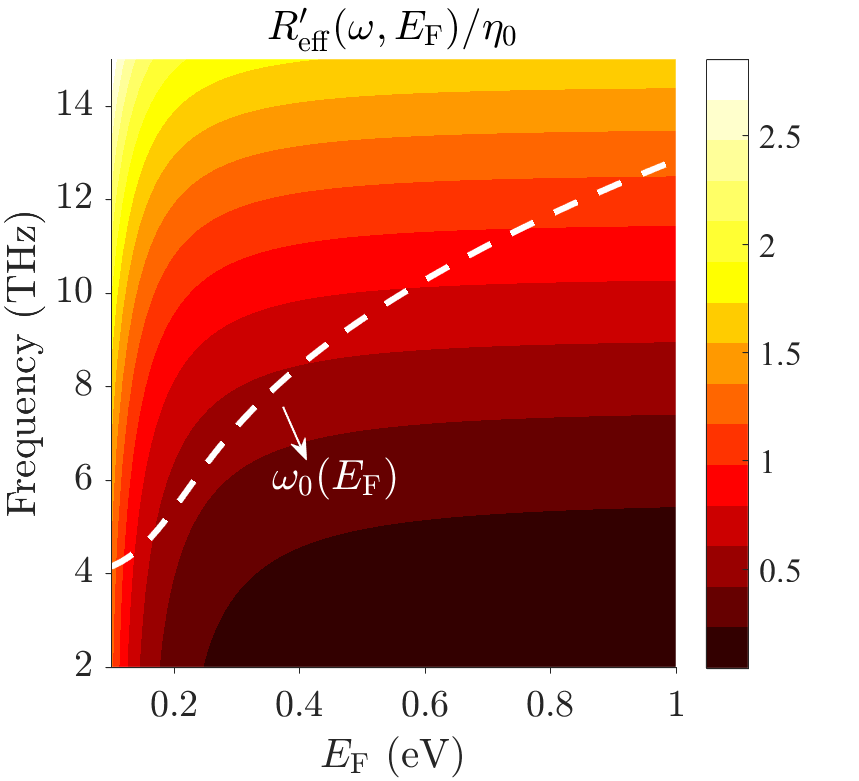}
		\label{fig:1500_mue_freq}}
	\subfigure[]{
		\includegraphics[width=0.3\linewidth]{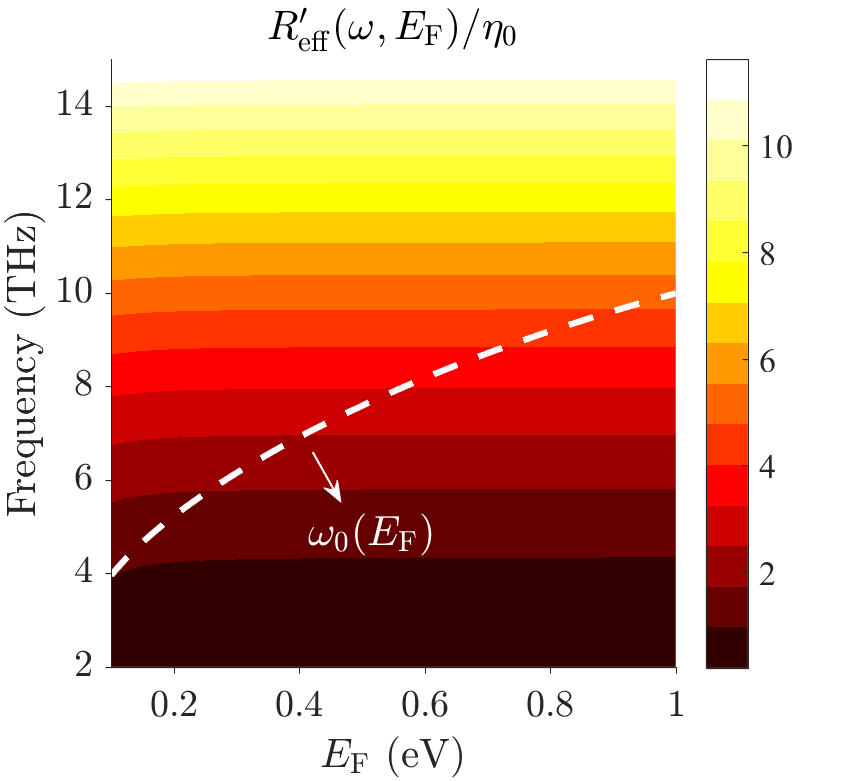}
		\label{fig:10000_mue_freq}}
	\caption{Normalized effective shunt resistance of graphene as a function of the frequency and Fermi level,  for graphene mobility (a) $\mu_{\rm m}=200~{\rm cm}^{2}{\rm V}^{-1}{\rm s}^{-1}$, (b) $\mu_{\rm m}=1500~{\rm cm}^{2}{\rm V}^{-1}{\rm s}^{-1}$ and (c) $\mu_{\rm m}=10000~{\rm cm}^{2}{\rm V}^{-1}{\rm s}^{-1}$.}\label{effective shunt resistance} \label{fig:mue_freq_R_eff}
\end{figure*}

In this section, we discuss tunability of  perfect absorption at THz frequencies as the Fermi level of graphene changes. 
To better explain the tunability property, first we simplify the complex Kubo formula of graphene conductivity in (\ref{Kubo}) into Drude model (valid in low THz band), $\sigma_{\rm g}={e^2E_{\rm F}}/[{\pi\hbar^2(j\omega+\gamma)}]$
which corresponds to the following expression for the sheet impedance of graphene:
\begin{equation}
	Z_{\rm g}=R_{\rm g}+j\omega L_{\rm g}=\frac{\pi\hbar^2v^2_{\rm F}}{e\mu_{\rm m} E^2_{\rm F}}+j\omega\left(\frac{\pi\hbar^2}{e^2E_{\rm F}}\right) . \label{drude_impedance}
\end{equation}
The effective shunt resistance and inductance of graphene  [see Fig.~\ref{fig:proposed_circuit_model}] can then be expressed in terms of $R_{\rm g}$ and $L_{\rm g}$: $	R_{\rm eff}^\prime(\omega, E_{\rm F})={R_{\rm g}}/{p}+{\omega^2 L^2_{\rm g}}/{R_{\rm g}p}$ and $L_{\rm eff}^\prime(\omega, E_{\rm F})={L_{\rm g}}/{p}+{R^2_{\rm g}}/{\omega^2 L_{\rm g}p}$.
Note that for a fixed value of graphene mobility, both $R_{\rm eff}^\prime$ and $L_{\rm eff}^\prime$ depend on the frequency and the Fermi level in graphene. The $E_{\rm F}$-dependent effective inductance indicates that the resonant frequency of the $RCL$-circuit in Fig.~\ref{fig:proposed_circuit_model} is tunable with the chemical potential. The resonant frequency $\omega_0$ can be solved from condition (\ref{condition: high impedance surface}):

\begin{equation}
	\frac{\omega_0 L_{\rm g}p}{\omega_0^2L^2_{\rm g}+R^2_{\rm g}}-\omega_0 C_{\rm m}-\frac{1}{\eta_{\rm d} \tan(\omega_0\sqrt{\epsilon_{\rm r}\epsilon_{\rm 0}\mu_{\rm 0}}d)}=0. \label{resonant_frequency}
\end{equation}

The curves in Fig. \ref{fig:mue_freq_R_eff} shows the resonant frequency of four different structures which all realize perfect absorption at $f_0=4$~THz and $E_{\rm F0}=0.1$~eV using graphene sheets of different  mobilities. The structural parameters are given  in the caption of Fig.~\ref{fig:differnt mobilities}. 
We can see that the resonant frequency always climbs up when the graphene Fermi level increases.
However, the modulation of $\omega_0$ does not necessarily mean frequency-tunable absorption. Another necessary condition is impedance matching with free space at the resonant frequency.
In order to realize frequency-tunable perfect absorption, the best scenario is to ensure that the condition $R_{\rm eff}^\prime(\omega_0, E_{\rm F})=\eta_0$ is always satisfied under the change of the  Fermi level.
Although the effective resistance of graphene is easy to tailor to $\eta_0$ at a fixed frequency and a given graphene state, it is hard to sustain this established impedance matching when the Fermi level of graphene is changing: The impedance matching condition is almost determined by the intrinsic properties of graphene.

Figure~\ref{fig:mue_freq_R_eff} shows the dependence of $R_{\rm eff}^\prime(\omega, E_{\rm F})$ on the frequency and the Fermi level for different qualities of graphene.
From Fig.~\ref{fig:200_mue_freq} we can see that for low-mobility graphene $R_{\rm eff}^\prime$ is quite susceptible to the Fermi level while it is not sensitive to the frequency. The value of $R_{\rm eff}^\prime/\eta_0$ corresponding to the resonant frequencies [the curve of $\omega_0 (E_{\rm F})$ in Fig.~\ref{fig:200_mue_freq}] decreases from unity to almost zero. Apparently,  very poor quality graphene is not suitable to achieve frequency-tunable perfect absorption. On the other hand, quickly decreasing  shunt resistance at a fixed frequency makes it a good candidate to realize a switch, which will be discussed in Section~\ref{Intensity-tunable absorber (Switch)}. For the graphene sheet with $\mu_{\rm m}=1500~{\rm cm}^{2}{\rm V}^{-1}{\rm s}^{-1}$, as the chemical potential increases, the frequency point where $R_{\rm eff}^\prime (\omega, E_{\rm F})=\eta_0$ also shifts to higher frequencies. This is in harmony  with increased resonant frequency. Also, the values of $R_{\rm eff}^\prime$ at the resonant frequencies are very close to $\eta_0$, which means that the peak absorption level remains close to unity even when the resonant frequency is tuned to higher values. For high-quality graphene ($\mu_{\rm m}=10000~{\rm cm}^{2}{\rm V}^{-1}{\rm s}^{-1}$), the effective resistance of graphene appears to be insensitive to the change of the  graphene Fermi level,  and increases gradually with the frequency. Therefore, it can be expected that the peak absorbance decreases with the increase of the chemical potential.

The above analysis suggests possibilities to realize broad tunability of graphene perfect absorber with both low and high quality graphene. In the following, we will show two different modulation scenarios in detail: frequency-tunable perfect absorbers and intensity-tunable perfect absorbers (switches) using low mobility graphene. 

\subsection {Frequency-tunable absorber} 

\begin{figure}[h!]
	\centering
	\includegraphics[width=0.9\linewidth]{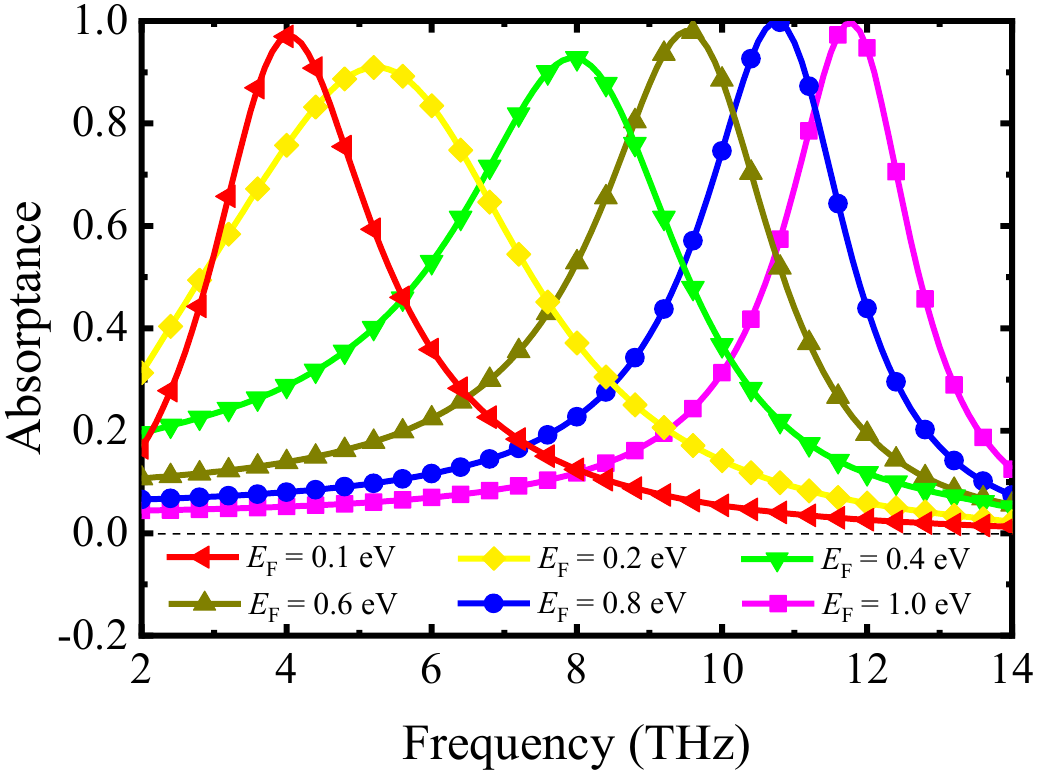}
	\caption{Simulated absorption intensity in terms of the frequency and doping level when varying the Fermi level from 0.1~eV to 1~eV. The graphene mobility is set to $\mu_{\rm m}=1500~{\rm cm}^{2}{\rm V}^{-1}{\rm s}^{-1}$ ($\tau=15$~fs at $E_{\rm F}=$~0.1~eV). }\label{fig:Frequency_tunable_simulaiton}
\end{figure}

Here we present tuning characteristics of a frequency-reconfigurable absorber designed based on the developed theory. The mobility of graphene is chosen as $\mu_{\rm m}=1500~{\rm cm}^{2}{\rm V}^{-1}{\rm s}^{-1}$, and the primary doping is assumed as $E_{\rm F0}=0.1$~eV. The structural parameters are given in the caption of Fig.~\ref{fig:differnt mobilities}.
Figure~\ref{fig:Frequency_tunable_simulaiton} shows the analytically calculated absorption coefficient from 1~THz to 14~THz when the Fermi level is raised from the original 0.1~eV to 1~eV. Apparently, the absorption frequency is blue-shifted from 4~THz to 12~THz, realizing a large modulation bandwidth (about 100\%). The peak absorbance at each Fermi level is very close to unity, only with a small degradation  around $E_{\rm F}=0.2$~eV (92\% absorption). 
The near perfect tunability in absorption frequency indicates that the critical coupling between light and graphene is sustainable during the Fermi level modulation, even if graphene's property dramatically changes.  
It should be pointed out that the mobility of graphene is not strictly limited at $\mu_{\rm m}=1500~{\rm cm}^{2}{\rm V}^{-1}{\rm s}^{-1}$. The most effective frequency tuning is achieved with the mobility between $\mu_{\rm m}=1000~{\rm cm}^{2}{\rm V}^{-1}{\rm s}^{-1}$ and $\mu_{\rm m}=2000~{\rm cm}^{2}{\rm V}^{-1}{\rm s}^{-1}$. This mobility range is also quite reasonable and can be obtained using CVD-grown graphene \cite{yao2013wide,jadidi2015tunable,ju2011graphene,ying2017enhanced}.


\subsection {Intensity-tunable absorber}  \label{Intensity-tunable absorber (Switch)}

As explained above, the rapidly decreasing $R^\prime_{\rm eff}$ of biased low-quality graphene makes it possible to be an efficient switchable absorber. Here we assume a graphene sample with a very poor mobility ($\mu_{\rm m}=200~{\rm cm}^{2}{\rm V}^{-1}{\rm s}^{-1}$) and study the tunability of the absorption level at the resonant frequency. Following the design procedures, the structural parameters are determined as $D=7~\mu$m, $p=137$ and $a=0.5$ if the perfect absorption is expected at 4~THz for $E_{\rm F0}=0.1$ eV. 


We note that,  in order to reduce the high resistance of such low-quality graphene, a narrow gap ($p=136$ and $g=50$~nm) between the metallic patches is required in this case.  This significantly increases the fabrication difficulties. The situation will become even worse if the operating frequency is higher ($D$ becomes smaller). On the other hand, the tiny width of metallic slots is not desirable, since the Fermi level pinning effect in graphene is not negligible. This practical issue is caused by the transfer of electrons in metal to the graphene sheet via the metal-graphene contact. The transferred charges result in an inhomogeneous distribution of the Fermi level in graphene \cite{khomyakov2010nonlinear,jadidi2015tunable}. 

\begin{figure}[h!]
	\centering
	\includegraphics[width=0.9\linewidth]{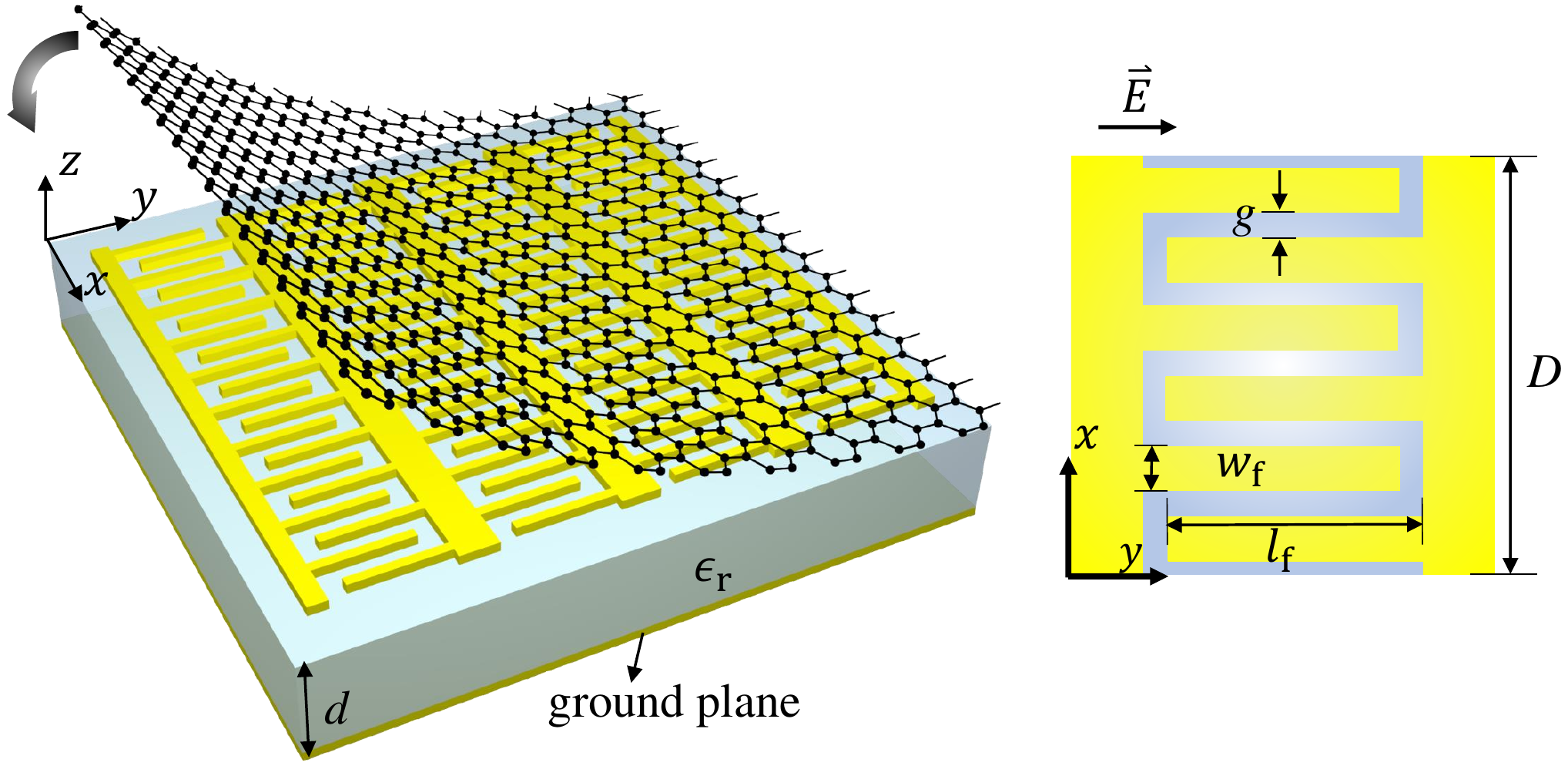}
	\caption{Schematic of the metasurface pattern with meandered slots (one square unit). The electric field is along the \textquotedblleft finger\textquotedblright~direction. This structure is appropriate  only for $y$-polarized incident wave. }\label{fig:meandered_structure}
\end{figure}

In order to avoid these problems, we propose a novel metallic pattern with meandered metal channels instead of straight ones (formed by the patch array), as shown in Fig.~\ref{fig:meandered_structure}. The $y$-oriented metal strips (called  \textquotedblleft fingers\textquotedblright~) intersect with each other to form a strong capacitance. The enhanced electric flux between the fingers dramatically increases the absorption in graphene. 
From the perspective of impedance matching, the meandered structure increases the length-width ratio of graphene, so that we can effectively reduce the shunt resistance of graphene sheet while keeping the gap relatively large. The required relations between the finger length, unit and gap sizes can be approximated as ${(Nl_{\rm f}+D)}/{g}=p$, where $N$ is the numbers of fingers in one square unit cell.

\begin{figure}[h!]
	\centering
	\includegraphics[width=0.9\linewidth]{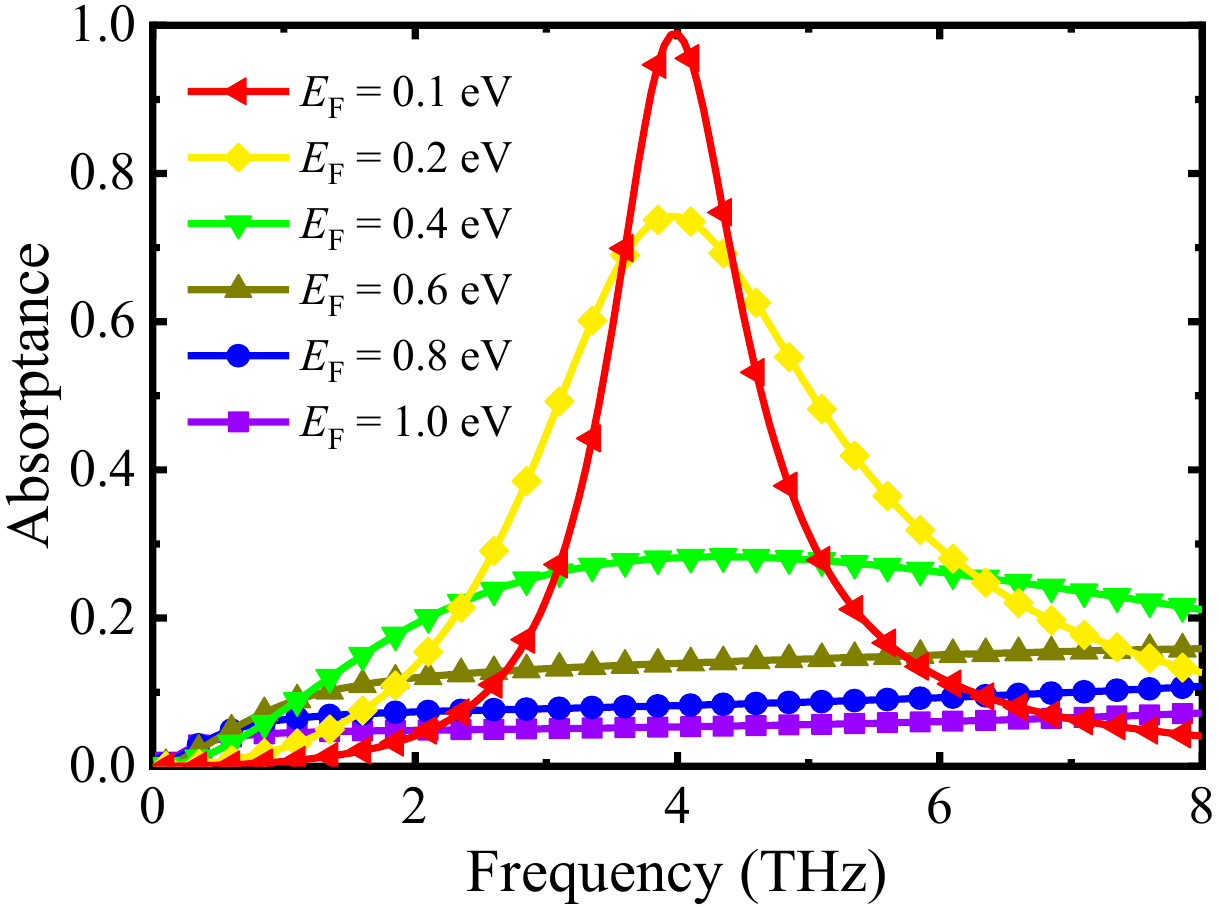}
	\caption{Absorption spectrum of the proposed meandered structure. The perfect absorption is optimized at 4~THz when $E_{\rm F0}=0.1$~eV and $\mu_{\rm m}=$ 200~${\rm cm}^{2}{\rm V}^{-1}{\rm s}^{-1}$ ($\tau=2$~fs). The substrate is chosen as $\epsilon_{\rm r}=2.8$ and $d=2.1~\mu$m. }\label{fig:tunable_intensity}
\end{figure}

Using graphene of the same quality ($\mu_{\rm m}=200~{\rm cm}^{2}{\rm V}^{-1}{\rm s}^{-1}$), perfect absorption is achieved when $D=5~\mu$m, $l_{\rm f}=3.6~\mu$m,   $g=250~$nm and $N=8$. Perfect absorbance is shown in Fig.~\ref{fig:tunable_intensity} (red curve). When the Fermi level is tuned from 0.1~eV to 1~eV, the absorption level  fastly decreases from unity to 5\%. The device acts as a perfect absorber in its natural state ($E_{\rm F0}=0.1$~eV) while in the biased state it is a nearly perfect reflector.

This design is very useful for perfect absorption in highly resistive thin films. For example, an undoped graphene sheet has the minimum conductivity (limit value) \cite{novoselov2005two}, $\sigma_{\rm g}^{\rm min}=e^2/h$ (or $R_{\rm g}^{\rm max}=25900$~ohm/sq). In the low terahetz band, where $R_{\rm g}^\prime\approx R_{\rm g}$, the scaling factor approximates to $p\approx R_{\rm max}/\eta_0=68$, which can be very easily realized with the meandered slots design. It should be noted that, in practice, there is always some background doping in graphene induced by many factors such as impurities, chemicals, and the substrate, which makes it difficult to reach that minimum conductivity limit.

\section{Experimental feasibility} 

\begin{figure}[h!]
	\centering
	\includegraphics[width=0.7\linewidth]{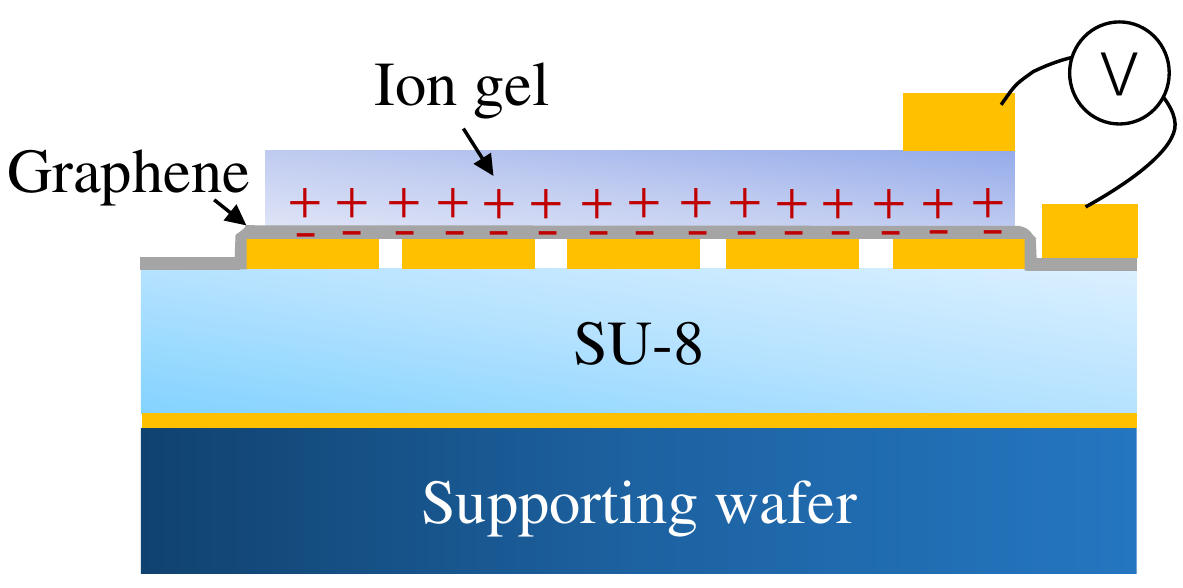}
	\caption{Fabrication schematic of tunable graphene absorbers based on metasurface substrates.}\label{fig:experimet method}
\end{figure}

Known experimental realizations of efficient graphene tunable devices are restricted by the low quality of graphene samples, as mentioned in the introduction section. Our proposed design method circumvents this limitation and allows strongly tunable absorption with arbitrary graphene mobility. Therefore, our designs are quite promising to be implemented in practice, for instance as the future graphene-based modulators. Figure~\ref{fig:experimet method} shows the scheme of possible device fabrication. The dielectric substrate is SU-8 ($\epsilon_{\rm r}\approx2.8$) which has been widely used in MEMS technology \cite{lorenz19978}. The thickness of SU-8 can be arbitrarily controlled by the spin coating speed. The gel-like SU-8 is then sintered in a high temperature environment, to be developed into a low-loss dielectric layer. Depending on the required line resolution, either e-beam or photo-lithography can be used to define the metallic patterns. Then a CVD-grown graphene sheet is directly transferred onto the fabricated metasurface substrate. One can choose ion gel as the gate dielectric to electrostatically dope graphene. Ion gel is well known for its high capacitance nature. 
Under a biased voltage between graphene and ion gel layer,  positive ions will concentrate in the vicinity of the graphene-gel interface (several nanometer away from the graphene plane), thus forming a high-capacitance gate. With this method, the Fermi level can be easily raised up to 1 eV only with a weak control voltage (less than 5~V), as demonstrated in many works \cite{safaei2017dynamically,kakenov2016observation,hu2015broadly}. The fabrication, electrical gating and measurements of the device can follow the methodology outlined in \cite{jadidi2015tunable} but with completely different functionalities.

\section{Applicability to other two-dimensional materials}
{Similarly with graphene, many other two-dimensional (2D) materials, such as semiconducting transition metal dichalcogenides (TMDs) \cite{docherty2014ultrafast} and black phosphorus (BP) \cite{tran2014layer}, also suffer from weak light-matter interactions.  According to the explanations in Section~\ref{section:limitation}, weak absorption stems from the impedance mismatch between the shunt sheet resistance of the material layers and the free-space impedance. In the case of larger shunt resistance than the free-space impedance, it is possible to use our proposed metasurface as a substrate for realizing perfect absorption. }

{Here, we consider an  example of monolayer black phosphorous which exhibits anisotropic Drude-type conductivity in the crystal plane, $\sigma_{\rm bp}^{ii}=-jn_{\rm c}e^2/m_i(\omega-j\gamma)$. In this expression,   $i$ labels the direction along $x$- (armchair) or $y$- (zigzag) axes,  $n_{\rm c}$ is the carrier density, and $m_i$ is the effective electron mass which is different along two orthogonal axes (the $x$- and $y$-directions) \cite{low2014tunable,liu2016localized}. According to \cite{low2014tunable}, the material parameters can be typically assumed as $\gamma=10$~meV, $n_{\rm c}=2\times10^{13}~{\rm cm}^{-2}$, $m_x=0.15m_0$ and $m_y=0.7m_0$ ($m_0$ is the static mass of electron). From 4~THz to 15~THz, the shunt resistance along  the $x$-direction  varies from 1500 to 16000~$\Omega$ which is much larger than $\eta_0$. This means that in this range we can use a patch-type metasurface to fully absorb $x$-polarized waves. All of the analytical formulas developed for graphene are valid for other 2D materials only by replacing the graphene impedance $Z_{\rm g}$ with the sheet impedance of thin layers  of other materials. 
The conductivity of other 2D materials is also tunable either by electrostatic or optical means. Thus, the absorption level can be dynamically modulated. The tunability analysis can follow the example of graphene discussed in Section~\ref{tunability}.}

\section{Conclusion}
To summarize,  we  have developed an analytical model of graphene-metasurface structures which allows us to identify physical mechanisms behind tunable absorption in graphene and design structures possessing required electromagnetic properties with the use of different graphene samples.
	Our systematic analysis demonstrated that, in the terahertz frequency range, perfect absorption as well as excellent tunability can be realized with graphene exempt from mobility requirements.
	In particular, using low quality graphene, we show that the absorption can be dynamically tuned either in the absorption frequency or absorption levels, with nearly 100\% modulation efficiencies. 
	The examples are provided for the terahertz frequency range, 
	but the developed graphene-metasurface analytical model is not limited to the terahertz band. It can be scaled from microwave to mid-infrared ranges by tailoring the metasurface dimensions. This theory is  applicable also to other two-dimensional materials such as black phosphorus and transition metal dichalcogenides.

\section*{Acknowledgments}
The authors would like to thank Zhipei Sun, Yun-Yun Dai, George Deligeorgis, Anna C. Tasolamprou, Odysseas Tsilipakos, Igor Nefedov, Fu Liu, Mohammad Sajjad Mirmoosa and Ana~D\'{i}az-Rubio for helpful technical discussions. This work was supported by the European Union's
Horizon 2020 Future Emerging Technologies call
(FETOPEN-RIA) under Grant Agreement No. 736876
(project VISORSURF).

\bibliographystyle{ieeetr}
\bibliography{graphene_TAP}
\bibliographystyle{ieeetr}
\bibliographystyle{unsrt}

\end{document}